\documentclass[intlimits,twoside,a4paper]{article}
\usepackage{amsmath,amsfonts,bm,amssymb}
\usepackage[T2A]{fontenc}
\usepackage[cp1251]{inputenc}

\usepackage{cmpj2}

\issue{2013}{16}{4}{43004}
\doinumber{10.5488/CMP.16.43004}

\title[Self-assembly of LC macromolecules]
{Relation between the grafting density of liquid crystal macromolecule
and the symmetry of self-assembled bulk phase: coarse-grained molecular
dynamics study}

\author{J.M.~Ilnytskyi}

\address{
Institute for Condensed Matter Physics
of the National Academy of Sciences of Ukraine,\\
1 Svientsitskii St., 79011 Lviv, Ukraine
}

\date{Received August 20, 2013}

\newcommand{\idx}[1]{_{\mathrm{#1}}}
\newcommand{\upx}[1]{^{\mathrm{#1}}}
\newcommand{\Ang}{\textrm{~\AA}}
\newcommand{\fs}{\textrm{~fs}}

\newcommand{\ns}{\textrm{ns}}
\newcommand{\mx}{\textrm{max}}

\newcommand{\atm}{\,\textrm{atm}}

\newcommand{\Nch}{N_{\mathrm{ch}}}

\begin{document}

\maketitle

\begin{abstract}
I consider a generic coarse-grained model suitable for the
study of bulk self-assembly of liquid crystal (LC) macromolecules.
The cases include LC dendrimers, gold nanoparticles modified
by polymer chains with terminating LC groups and others. The study
is focused on the relation between a number of grafted chains,
$\Nch$, and the symmetry of the self-assembled bulk phases. Simple
space-filling arguments are used first to estimate stability
intervals for a rod-like, disc-like and spherulitic conformations
in terms of $\Nch$. These are followed by coarse-grained molecular
dynamics simulations for both spontaneous and aided self-assembly
of LC macromolecules into bulk phases. In spontaneous self-assembly
runs, essential coexistence of rod-like and disc-like conformations
is observed (via analysis of the histograms for the molecular
asphericity) in a broad interval of $\Nch$, which prevents the
formation of defect-free structures. The use of uniaxial and planar
aiding fields is found to improve self-assembly into monodomain phases
by promoting conformations of respective symmetry. Strong shape-phase
relation, observed experimentally, is also indicated by the simulations
by the coincidence of the stability intervals for the respective
conformations with those for the bulk phases.

\pacs 02.70.Ns, 61.30.Vx, 61.30.Cz, 61.30.Gd
\end{abstract}

\section{Introduction}\label{I}

When different (in size, shape or interaction potential) molecular
fragments are combined into a single molecule, then one obtains the
so-called polyphilic material \cite{Riess2003,Olsen2008} (the simplest example
being well-known amphiphiles). If such a molecule is big enough, then it represents
a supermolecular object, although the supramolecular effects are also
possible via self-assembly of polyphilic macromolecules into bulk ordered
phases \cite{Zeng1997,Discher2002,Gittins2003,Percec2004}. The self-assembly
is predominantly driven by a microphase separation, which, in turn,
depends on both the details of molecular architecture and on the level of chemical
compatibility between the constituent parts of a molecule.

LC polyphilic macromolecules incorporate mesogenic groups
in addition to branched polymer chains, nanoparticles, etc. Variation of a
molecular architecture gives rise to main- or side-chain LC polymers, LC
dendrimers and elastomers, LC gold metamaterials
\cite{Tsc01b,SG05,Tschierske2007,Saez2008}.
The microphase separation in such systems originates from poor
miscibility of aromatic and aliphatic fragments, and/or size differences between
them (e.g., larger nanoparticle and smaller polymer bead or mesogen) as well as
on the other details of interparticle interactions.

Despite a broad variety of possible molecular architectures, certain cases
bear prominent similarities. Good example is provided by the existence of many
common features in self-assembly
of LC dendrimers and LC gold metamaterials \cite{SG05,Saez2008,Draper2011,Kumar2011},
and I will concentrate on these particular cases in this study.
Both systems exhibit a similar set of lamellar, columnar and various cubic phases
\cite{SG05,Saez2008,Draper2011,Kumar2011,PBS+00,Agina2007,Wojcik2009,Wojcik2010,Wojcik2011}.
Strictly speaking, the interior of these types of macromolecules is rather different:
flexible hyperbranched polymer scaffold {\it vs} solid nanoparticle, respectively.
This implies a different type of grafting in each case: more of annealed type
for LC dendrimer (the level of rearrangement freedom for grafted beads depends
on the dendrimer generation \cite{PBS+00,Agina2007}), and more of quenched
type for gold metaparticle. Nevertheless, the general aspects of a self-assembly
turned out to be more dependent on the space filling capabilities of grafted chains and
on the strength of the mesogen-mesogen interaction \cite{SG05,Draper2011,Kumar2011}
than on the details of internal structure of the macromolecule.

The similarities between the LC dendrimers and LC gold metaparticles open
up a possibility to describe their self-assembly by some generic
coarse-grained model, in which less relevant internal degrees of freedom
of a central core are neglected and only rearrangement of the attached polymer
chains with terminating mesogens is taken into account. The grounds
for such coarse-graining (besides general arguments given above) are
to be found in some previous simulation studies
\cite{WIS03,Balabaev_old,Balabaev_new}, where, in particular, it was found that
the central core of the generation three carbosilane dendrimer is on
average spherically symmetric in all (isotropic, nematic and smectic
A) phases of LC solvent \cite{WIS03}. The models that exploit this
fact have already been considered, namely in the form of a sphere with
attached chains, each containing a mesogen \cite{HWS05,ILW2010},
as well as a sphere decorated by Gay-Berne particles directly on its surface
\cite{OZ2013}. The number of bulk phases have been found in these
simulation works. The models, however, permit tuning in a number of ways
(the number and length of grafted chains, the precise way of grafting, the way
terminal mesogens are attached, etc.) and the effects of all these changes
still await to be analysed in detail by computer simulations.

The experimental studies reveal the existence of strong dependence
between the density of mesogens on the macromolecule surface,
molecular conformation in bulk phase and the symmetry of the latter
\cite{SG05,Saez2008,Draper2011,Kumar2011,PBS+00,Agina2007,Wojcik2009,Wojcik2010,Wojcik2011}.
As remarked in reference \cite{SG05}, the grafting density ``can effectively
change the overall gross shape of the structure of the supermolecule from
being rod-like, to disc-like, to spherulitic. Thus, the structure of the
systems at a molecular level can be considered as being
deformable, where each type of molecular shape will support
different types of self-organized mesophase structure. Thus, for
supermolecular materials, rod-like systems will support the
formation of calamitic mesophases (including various
possibilities of smectic polymorphism), disc-like systems tend
to support columnar mesophases, and spherulitic systems form
cubic phases''. The current study addresses this effect by means of
computer simulation.

The work is a continuation of the study performed in reference \cite{ILW2010},
where the coarse-grained generic model for the LC metaparticle was
introduced and studied on a subject of a bulk self-assembly. The model
contained a central sphere and 32 free-sliding chains each terminated
by a mesogen and the simulations were performed by means of a coarse-grained
molecular dynamics (CGMD). It has been found that the melt self-assemble
into either smectic lamellar or hexagonal columnar morphology
when aided briefly by an uniaxial or planar external field, respectively.
The molecular shape is predominantly rod-like in the smectic phase and
disc-like in the columnar one. This shape bistability was analysed by
means of average metric properties (gyration tensor, average asphericity,
etc.). On the contrary, the unaided (spontaneous) self-assembly by means
of either slow compression or cooling the melt down was found to
always result in the polydomain phase. Here, these findings are extended in two
directions. Firstly, the model is generalized to the case of an arbitrary
number of grafted chains $\Nch$, which allows one to study the intervals
of stability for each bulk phase on $\Nch$ by means of CGMD simulations.
Secondly, both aided and spontaneous self-assembly is analysed in detail
by splitting the melt into subsystems of rod-like and disc-like molecules
and monitoring the histograms of their asphericities. The phase boundaries
obtained by means of CGMD are also compared with the results of purely
geometric analysis for athermal space-filled rod, disc and sphere.

The following section contains a description of the model and a space-filling
analysis. Section \ref{III} contains the results for the CGMD simulations of
the bulk phases by means of spontaneous and aided self-assembly, as well as
a detailed analysis of molecular conformations. Conclusions are provided
in section~\ref{IV}.

\newpage
\section{Modelling and computational details}
\label{II}

\begin{figure}[ht]
\begin{center}
\includegraphics[clip,height=4cm]{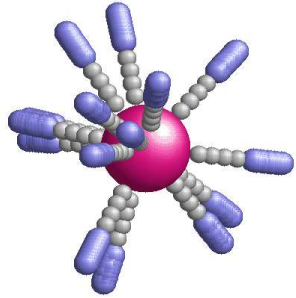}
\caption{\label{model_snap}(Color online) Generic coarse-grained model of liquid crystal
colloid consisting of large central sphere and $\Nch$ freely-grafted
chains each terminated by a mesogen.}
\end{center}
\end{figure}

The coarse-grained model for LC dendrimer or LC gold metaparticle (hereafter referred to
as ``generic model'') is depicted schematically in
figure~\ref{model_snap}. Large central sphere represents a  coarse-grained core
of a macromolecule with its internal degrees of freedom being neglected. Four
smaller spheres (each being a fragment of a polymer chain
of a few hydrocarbons) form a spacer. The latter is terminated by a
spherocylinder representing a coarse-grained mesogenic (LC) group.

This model, introduced in reference \cite{ILW2010} and studied there for the case of
$\Nch=32$ attached chains only, is generalised here for the case of arbitrary
$\Nch$. The first bead of each chain can be attached to the surface of a central
sphere in a number of ways, in particular:
(i) quenched-like grafting to a particular point on a surface;
(ii) semi-quenched-like grafting with the employment of an angular elastic spring with respect to a
particular point;
(iii) annealead-like grafting, when the end bead is capable of sliding freely on the surface.
In all cases, radial elastic spring can be used to ensure that the first bead is
 always located on the surface of a large sphere.
It is evident that option (i) would be the best suited to model the
LC gold metaparticle, whereas option (ii) would represent the LC dendrimer (e.g., such option
was applied in \cite{HWS05}). Option (iii) can be seen as some limit case representing
the infinite generation LC dendrimer or the metamolecule with an additional symmetry
of chains interexchange. The latter is not unreasonable for the equilibration
speed-up and is, in fact, on par with high interpenetrability of soft beads in CGMD
modelling employed here. The option (iii) with annealed grafting is used in this study.

The effective dimensions of soft beads are based on the coarse-graining of the atomistic
model for the generation 3 LC dendrimer \cite{HWS05}.
These are: $\sigma_1=21.37\Ang$ for a large
sphere, $\sigma_2=6.23\Ang$ for the first bead of a spacer, $\sigma_3=4.59\Ang$
for all the rest beads of the spacer and $D=3.74\Ang$, $L/D=3$ for the mesogen
breadth and elongation, respectively. These dimensions are also used for the
visualisation purpose. The interaction potential between any two spheres has a
quadratic form:
\begin{equation}\label{Uspsp}
V_{ij}\upx{sp-sp}=\left\{
\begin{array}{ll}
U_\mx\upx{sp-sp}(1-r^*_{ij})^2, & r^*_{ij}<1,\\
0,                              & r^*_{ij}\geqslant  1,
\end{array}
\right.
\end{equation}
where $r^*_{ij}=r_{ij}/\sigma_{ij}$ is the scaled distance between the
centers of $i$-th and $j$-th sphere, and usual mixing rules
$\sigma_{ij}=(\sigma_i+\sigma_j)/2$ are employed for the spheres with
different diameters $\sigma_i$ and $\sigma_j$. The value of
$U_\mx\upx{sp-sp}=70\cdot 10^{-20}$~J is the same for all combinations of interacting
spheres. The same potential form is used for interaction between
the sphere and the spherocylinder:
\begin{equation}\label{Uspsc}
V_{ij}\upx{sp-sc}=\left\{
\begin{array}{ll}
U_\mx\upx{sp-sc}(1-d^*_{ij})^2, & d^*_{ij}<1,\\
0,                              & d^*_{ij}\geqslant  1,
\end{array}
\right.
\end{equation}
where $d^*_{ij}=d_{ij}/\sigma_{ij}$ is a dimensionless closest distance
between the center of the $i$-th sphere and the core of the $j$-th
spherocylinder, with the scaling factor $\sigma_{ij}=(\sigma_i+D)/2$. Parameter
$U_\mx\upx{sp-sc}$ is equal to $U_\mx\upx{sp-sp}$ (see above).

Spherocylinder-spherocylinder pairwise interaction has the form introduced
by Lintuvuori and Wilson \cite{LW08}:
\begin{equation}\label{Uscsc}
V_{ij}\upx{sc-sc}=\left\{
\begin{array}{ll}
U_\mx\upx{sc-sc}(1-d^*_{ij})^2+\epsilon^*, & d^*_{ij}<1,\\
U_\mx\upx{sc-sc}(1-d^*_{ij})^2-U^*_{\mathrm{attr}}(\hat{r}_{ij},\hat{e}_i,\hat{e}_j)
     (1-d^*_{ij})^4+\epsilon^*, & 1 \leqslant  d^*_{ij}<d^*_{\mathrm{c}}\,,\\
0,                              & d^*_{ij} > d^*_{\mathrm{c}}\,,
\end{array}
\right.
\end{equation}
where $d^*_{ij}=d_{ij}/D$ is the dimensionless nearest distance
between the cores of spherocylinders \cite{EIW01}, $d^*_{\mathrm{c}}$ is the effective cutoff
distance for the attractive interaction that depends on the attractive part of
the potential
\begin{equation}\label{Uattr}
U^*\idx{attr}(\hat{r}_{ij},\hat{e}_i,\hat{e}_j) =
U^*\idx{attr} - \left[
5\epsilon_1 P_2(\hat{e}_i\cdot\hat{e}_j)+
5\epsilon_2
(P_2(\hat{r}_{ij}\cdot\hat{e}_i)+P_2(\hat{r}_{ij}\cdot\hat{e}_j)
\right].
\end{equation}
The latter depends on the orientations $\hat{e}_i$, $\hat{e}_j$ of the long
axes of spherocylinders and the unit vector $\hat{r}_{ij}$ that
connect their centers, as discussed in more detail elsewhere \cite{LW08}.
$P_2(x)=1/2(3x^2-1)$ is the second Legendre polynomial, the energy
parameters are as follows: $U_\mx\upx{sc-sc}=70\cdot 10^{-20}$~J,
$U^*\idx{attr}=1500\cdot 10^{-20}$~J, $\epsilon_1=120\cdot 10^{-20}$~J
and $\epsilon_2=-120\cdot 10^{-20}$~J. The phase diagram of the system
of LC particles interacting via this potential is discussed in
\cite{LW08}.

Bonded interactions include harmonic bond and harmonic pseudo-valent angle
(introduced to mimic spacer rigidity on a coarse-grained level)
contributions:
\begin{equation}
 V_{\mathrm{bonded}}=\sum_{i=1}^{N_{\mathrm{b}}}k_{\mathrm{b}}(l_i-l_0^k)^2 + \sum_{i=1}^{N_{\mathrm{a}}}k_{\mathrm{a}}(\theta_{i}-\theta_0)^2,
\end{equation}
where $l_i$ and $\theta_i$ are instant values for $i$th bond length and $i$th pseudo-valent
angle (defined between each three consecutive beads in a spacer), respectively,
$N_{\mathrm{b}}$ and $N_{\mathrm{a}}$ being their maximum numbers.
Force constants are: $k_{\mathrm{b}}=50\cdot 10^{-20}$~J/$\!\Ang^2$ and $k_{\mathrm{a}}=20\cdot 10^{-20}$~J/rad$^2$,
bond length constants are: $l_0^1=14.9\Ang$ (large sphere--first sphere of spacer),
$l_0^2=3.6\Ang$ (first--second sphere of spacer), $l_0^3=3.62\Ang$ (all other bonds
between spheres in the spacer), $l_0^4=2.98\Ang$ (last sphere of a spacer--mesogen nearest
cap center). The pseudo-valent angle constant is $\theta_0=\pi$.

\begin{figure}
\begin{center}
\includegraphics[clip,width=3.5cm]{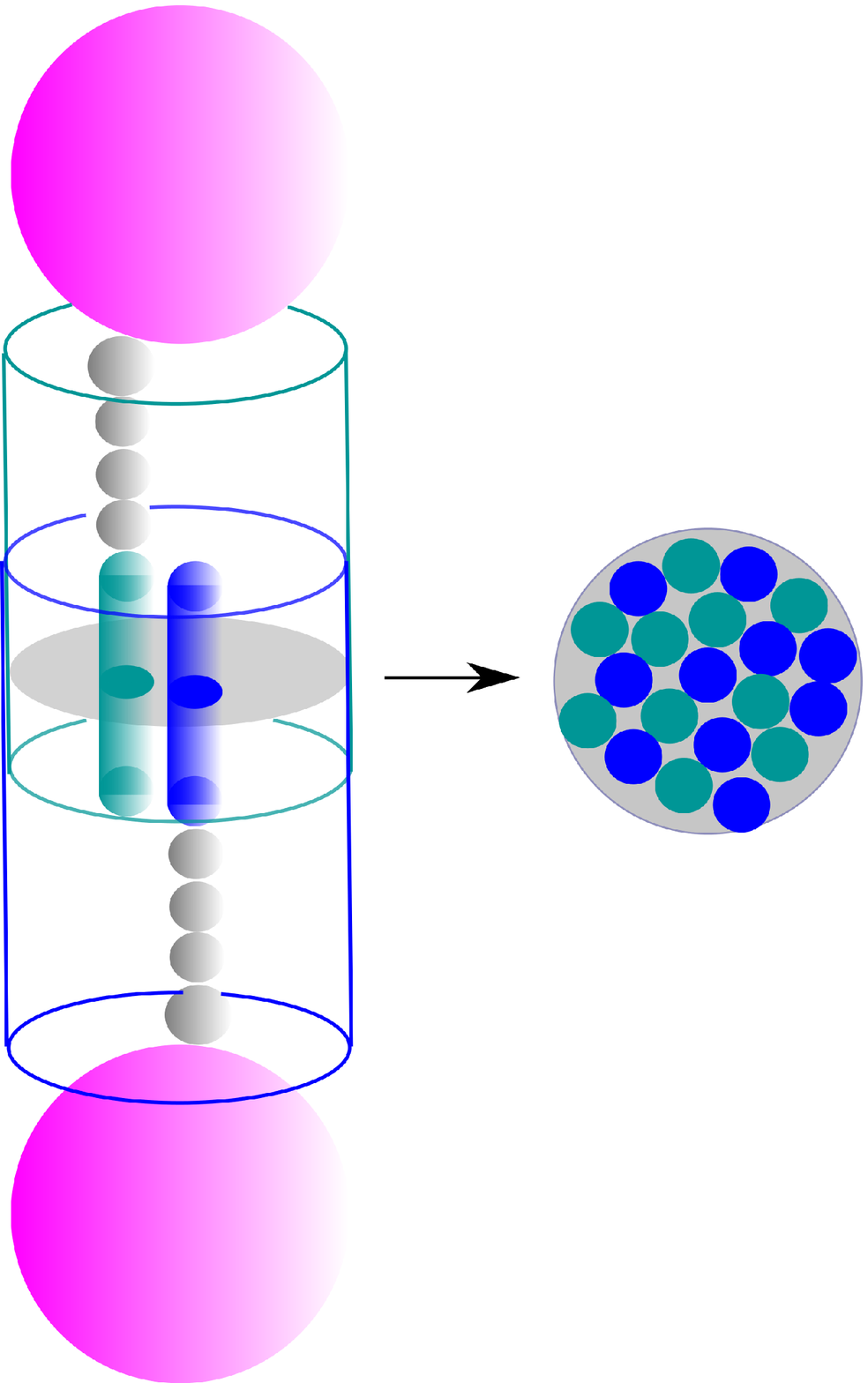}\hspace{2cm}
\includegraphics[clip,width=3.5cm]{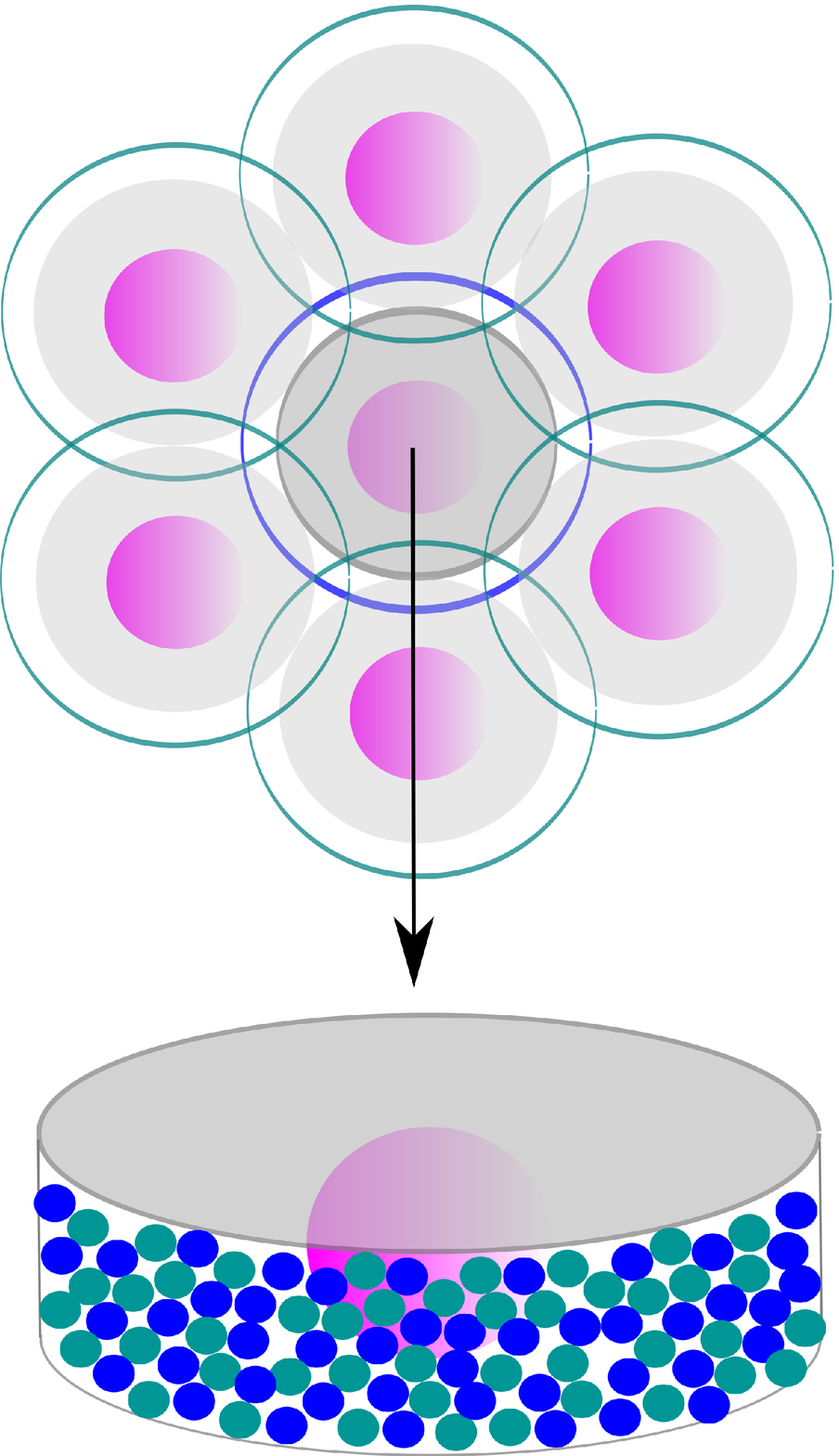}
\caption{\label{rod-disc-schema}(Color online) On  the left: cross-section region (shown as gray
circle) between the tails of two adjacent molecular rods packed in an interdigitated
smectic layer. The arrow points to 2D illustration of mesogens cross-sections packing
inside the cross-section region (blue and green discs represent mesogens from different
molecules). On the right: the same for molecular discs packed into interdigitated
hexagonal columnar phase. The side surface of discs (shown below) is the cross-section
region in this case, blue and green discs represent mesogens from the central and neighbouring
molecules, respectively.}
\end{center}
\end{figure}

Let me now consider possible conformations that can be observed in such model macromolecule
depending on a number of attached chains $\Nch$. Following experimental work
\cite{SG05,Saez2008,Draper2011,Kumar2011,PBS+00,Agina2007,Wojcik2009,Wojcik2010,Wojcik2011},
one would expect the possibility for the rod-like, disc-like and spherulitic shapes. It is obvious
that one of the crucial factors that will define the most favourable shape(s) at given
$\Nch$ is the capability of the available molecular elements of space-filling into a required form.
It is also known from both experimental \cite{SG05,Saez2008} and simulation \cite{ILW2010}
works that the mesogens of adjacent molecules highly interdigitate.
For the case of a rod-like conformation (in the smectic phase), one may consider
the ``slim rod'' limit when the breadth of the molecular rod is equal to the
diameter of the large sphere $\sigma_1$. If such two rods interdigitate, then the mesogens from both
molecules cross the mid-distance cross-section of diameter $\sigma_1$ (shown in grey in
figure~\ref{rod-disc-schema}, on the left). The condition of tight spacefilling of each molecule
into a rod is reduced then to close packing of 2D discs of diameter $D$ inside the circle of diameter
$\sigma_1$.
The number of hexagonally closely packed mesogens per cross-section
circle is $N'=k\frac{\pi r_1^2}{\pi(D/2)^2}=
k\left(\frac{r_1}{D/2}\right)^2\approx 30$, where $k=0.91$ is a packing
fraction for 2D hexagonal lattice and $r_1=\sigma_1/2$. Half of these (shown as blue)
belong to the lower molecule only, but each molecular rod has two tails.
Therefore, the number of chains per molecule in the ``slim rod'' limit is
$N_{\mathrm{rod}}=N'\approx 30$. This is an estimate for the average number of
chains to form a tightly space-filled rod.

Similar estimates can be performed for the case of a disc-like conformation
(in the columnar phase) in a ``slim disc'' limit (see, figure~\ref{rod-disc-schema},
on the right). In this case, the width of
the disc is equal to $\sigma_1=2r_1$ and its radius $R_{d}$ can be estimated from the
sums of bond lengths in the spacer and half a length of the mesogen,
yielding $R_{d}\approx 34.3\Ang$ (the half of the mesogen length is taken into
account due to mesogens interdigitation with those from six neighboring
molecules). The number of closely packed mesogens on the side surface of a disc is,
therefore, $N''=k\frac{2\pi R_{d}\cdot 2r_1}{\pi (D/2)^2}\approx 382$. Only half of these
mesogens belong to a given molecule (shown as blue discs in the cross-section
region in figure~\ref{rod-disc-schema}), hence the number of chains per molecule
in a ``slim disc'' limit is $N_{\mathrm{disc}}=N''/2\approx 191$. This number, however, turns out to be
unrealistic for our model, because one should take into account that the density of
chains increases closer to the central sphere. Indeed, the number of closely
packed grafting beads (of radius $r_2=\sigma_2/2$) attached to the side surface
of a small disc of a radius $r_1+r_2$
(made around a central sphere) is only $N^*=k\frac{2\pi (r_1+r_2)\cdot 2r_1}{\pi r_2^2}
\approx 55$, four times less than it is required for close packed
mesogens on the edge surface of a disc-like molecule. Therefore, at $\Nch>N^*$ one would face
a tremendous crowding of beads near the surface of a large sphere and the reasonable estimate
for $\Nch$ to form space-filled (near the central sphere only) disc would be
$N^*\approx 55$. For the case of spherulitic conformation, the
situation is similar and the close packed external shell cannot be achieved due to
limitations on the grafting density at the surface of a central sphere. The number of closely grafted
polymer beads in this case is estimated as $N^\dagger=k\frac{4\pi (r_1+r_2)^2}{\pi r_2^2}\approx 71$.

This analysis, based on space-filling of molecular elements, results in a very rough estimate
for the average number of chains $\Nch\sim 30, 55, 71$ that are optimal to form a rod-like,
disc-like and spherulitic space-filled conformations, respectively. It leaves beyond the
effect of conformational entropy, which results in swelling of both rods and discs, and this
will be temperature dependent. The equilibrium conformation (and the resulting bulk morphology)
will be the result of the competition between enthalpy of the mesogen-mesogen interactions
and various entropic contributions to the free energy. The effects are taken into account
most naturally in the CGMD simulations presented in the following section.

\section{Bulk phases, aided and spontaneous self-assembly, analysis of molecular conformations via CGMD simulations}
\label{III}

Here, I use the same coarse-grained MD approach as was used in reference \cite{ILW2010}. This
is a pretty standard MD technique only to be applied to the system with soft coarse-grained
potentials, the details can be found in references \cite{ILW2010,IW_2000,IW_2001}.
The number of macromolecules being simulated is $N_{\mathrm{mol}}=100$ for each case of
$\Nch=8-64$ grafted chains, the $NPT$ and $NP_{xx}P_{yy}P_{zz}T$ ensembles are
used at the pressure of $53\atm$, the timestep is $20\fs$ and the leap-frog integrator
is employed.

It is assumed that the generic model for LC macromolecule (introduced in the previous
section and shown in figure~\ref{model_snap}) is capable of self-assembling into the following
bulk phases: lamellar smectic (macromolecules adopt a rod-like conformation),
hexagonal columnar (macromolecules adopt a disc-like conformation) and cubic
phase of possibly various symmetries. As already mentioned above, the grounds for
this are to be found in both experimental \cite{SG05,Saez2008,Kumar2011,Wojcik2011}
and simulation \cite{ILW2010} studies.

\begin{figure}
\begin{center}
\includegraphics[clip,width=3.7cm]{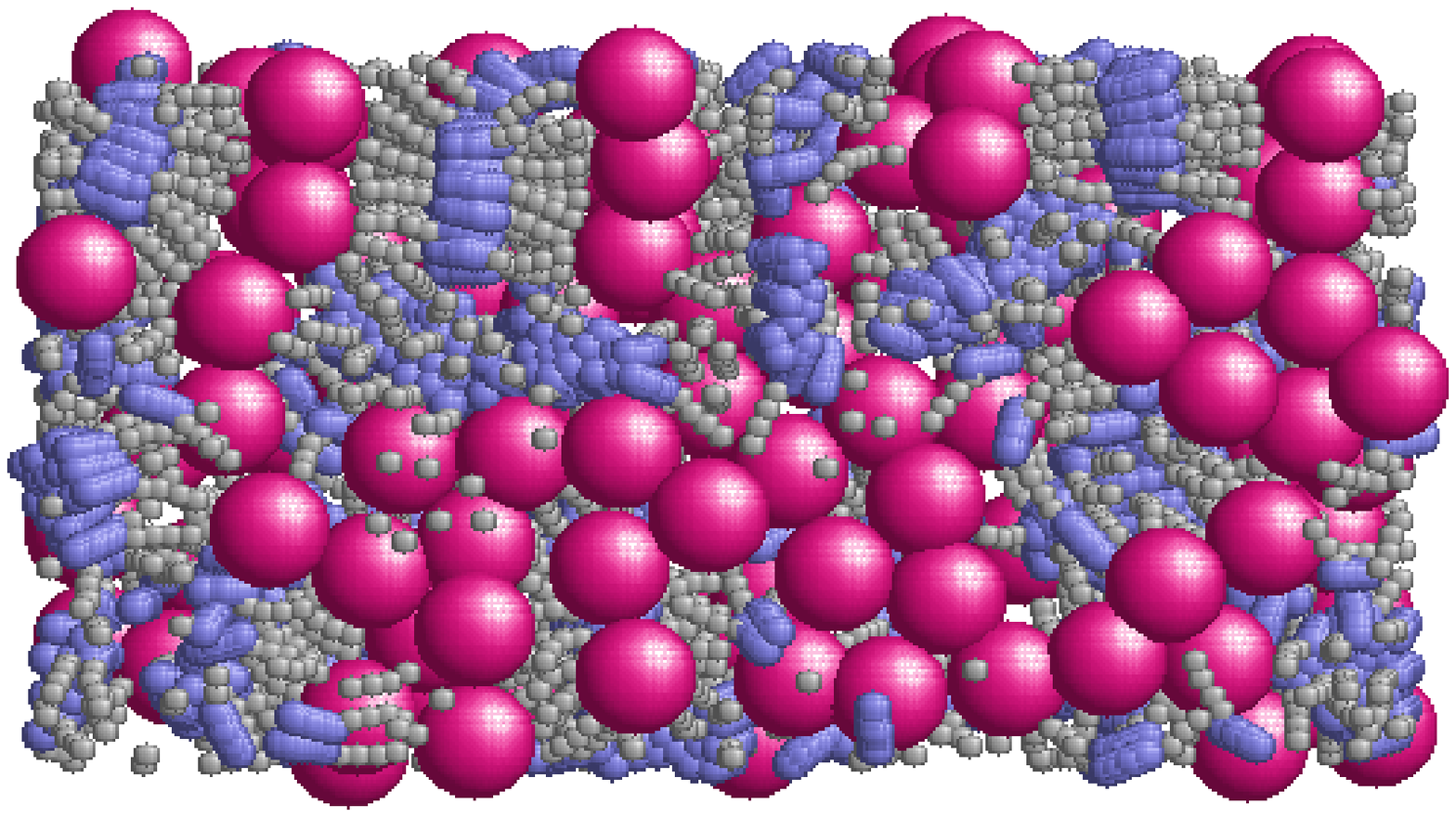}~~~~~~~~~~~~~~
\includegraphics[clip,width=5cm]{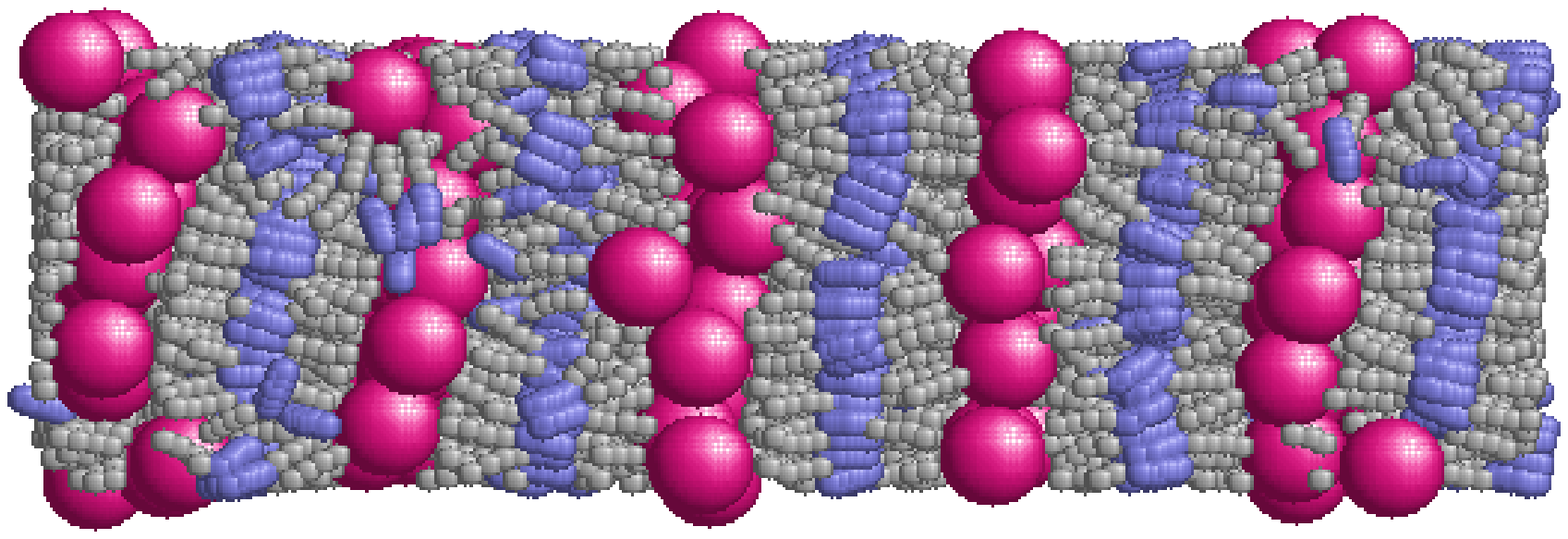}\\
\includegraphics[clip,width=5cm]{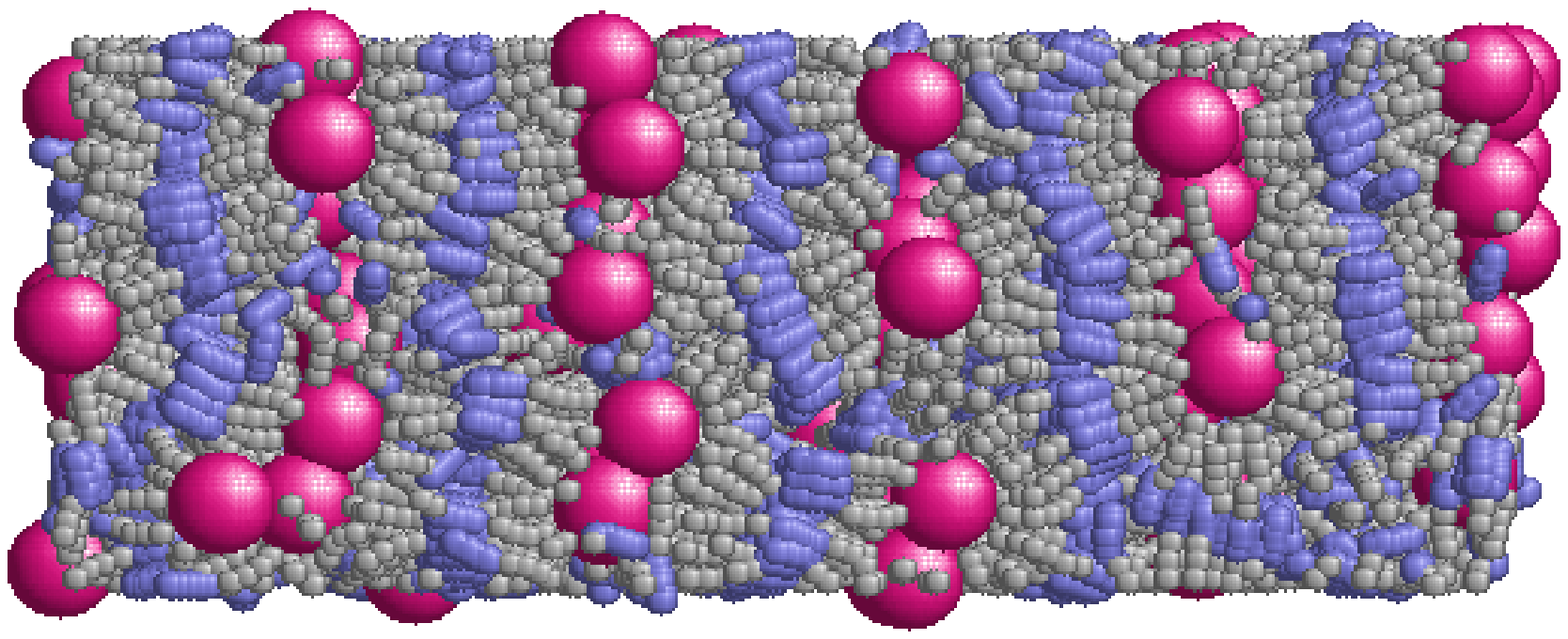}~~~
\includegraphics[clip,width=4.5cm]{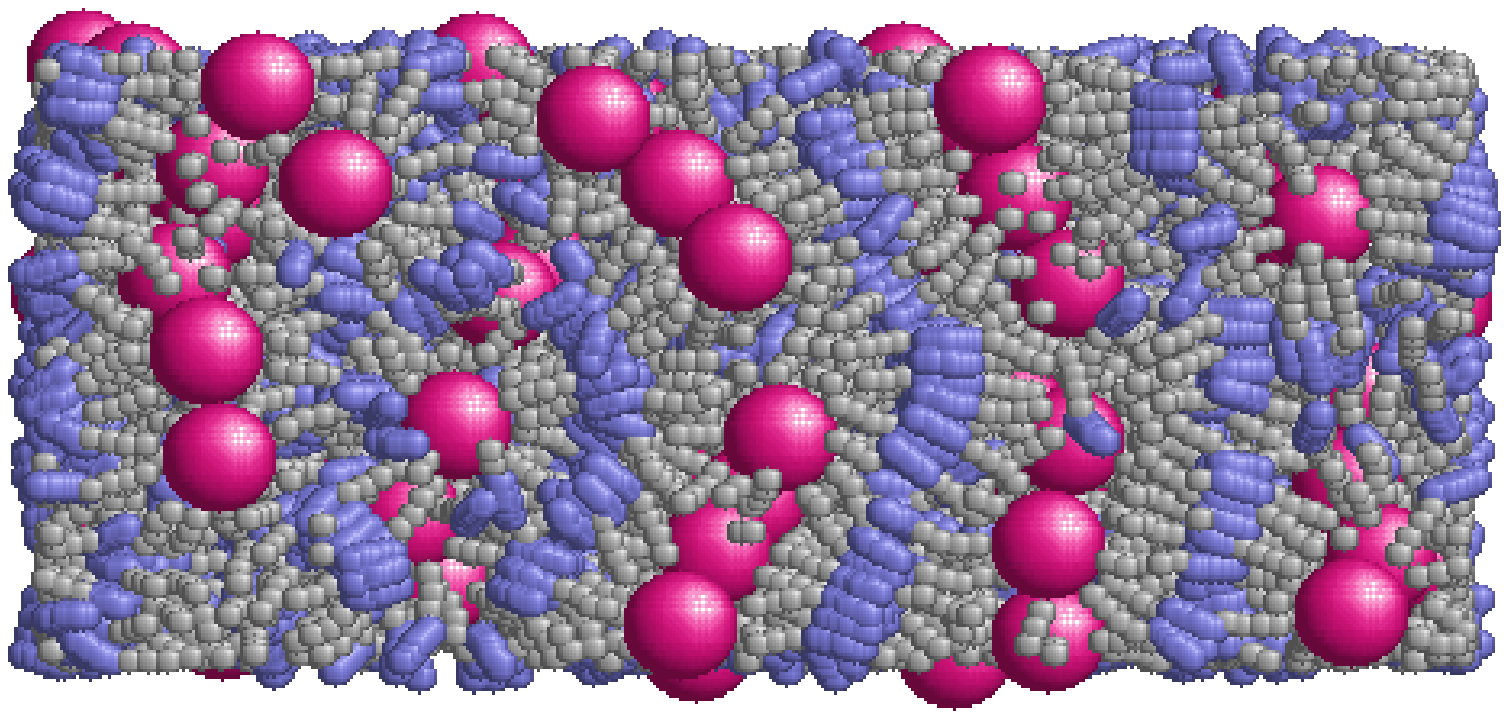}
\caption{\label{nch_f=0}(Color online) Snapshots for lamellar smectic phases obtained via
spontaneous self-assembly of generic model by cooling the sample from $T=500$~K
down to $450$~K with the cooling rate of $2.5$~K/$\ns$. Top left: $\Nch=8$, top right:
$\Nch=12$, bottom left: $\Nch=20$, bottom right: $\Nch=24$.}
\end{center}
\end{figure}

Slow self-assembly of LC macromolecular melts poses serious problems to computer
simulations. Essential speed-up for microphase separation can be achieved by using soft
potentials [e.g., equations~(\ref{Uspsp})--(\ref{Uscsc})], since in this case the beads are
semi-transparent and may overlap and cross each other during the
equilibration (see, e.g.,
\cite{HWS05,LW08,GM98,IlnPatsHol2008,Bates2009,Bates2009a,LW09}).
However, for the case of the model depicted in figure~\ref{model_snap}, the spontaneous
self-assembly was still found to typically lead to the polydomain (globally isotropic)
phase, both in the case of slow cooling down or slow compressing (the results
for $\Nch=32$ chains are discussed earlier \cite{ILW2010}).
Similarly to these findings, spontaneous self-assembly at a broader interval
of values of $\Nch=8-64$ turns out to be also more ``hit and miss''. I used relatively
slow cooling, when the temperature was lowered
linearly from $T=500$~K down to $450$~K during first $20~\ns$ (cooling rate is $2.5$~K/$\ns$),
followed by another run for $20~\ns$ at fixed $T=450$~K. As the result, relatively
defect-free smectic layers are found for the cases of $\Nch=12$ and $\Nch=20$,
whereas at other values of $\Nch\leqslant  24$, the polydomain layered structures have
been obtained (see, figure~\ref{nch_f=0}) with the sample preparation path being the
same in all cases.

There seem to be several reasons for hampering the spontaneous self-assembly of our model. The first one could be related to the annealed grafting of chains, which
results in a broad uncontrolled distribution of molecular asphericity (see below)
as well as may enhance microphase separation between large and small spheres, as
evidenced for the case of $\Nch=8$ (see, figure~\ref{nch_f=0}).
The second reason is high metastability of the melt below LC transition.
For instance, when the system is cooled down, once the
mesogens start to form LC domains, it is locked into a random network
formed by physical crosslinks between mesogens. As a result, the system is stuck
in a metastable state and cannot be driven further to the global minimum morphology
without applying a certain external stimulus. In real life, the perturbations of
various kind do exist, e.g., random flows (when melt is poured into some vessel),
centrifugal forces (when spin-coating is used), possibility to apply shear, laminar
flow or external fields. These stimuli constantly ``shake'' the molecules in various
ways and drive the melt towards the equilibrium state. Similar approaches could be
also used in MD simulations.

In reference \cite{ILW2010} the external fields acting on the mesogens were used to aid
the formation of bulk phases, this approach being also adopted in our study. The external
field is introduced via additional energy term:
\begin{equation}\label{urot}
 V^{\mathrm{rot}}_i=-F(\hat{e}_i\cdot\hat{f})^2,
\end{equation}
where $F$ is the amplitude of the field (the reduced amplitude $f$ will be defined as
$F=f\cdot 10^{-20}$~J), $\hat{e}_i$ is the unit vector directed along the long axis of
$i$th mesogen and $\hat{f}$ is the unit vector that defines the direction of the field.
When $F>0$, the field has an uniaxial symmetry, when $F<0$, its symmetry is planar
(promoting the orientation of the mesogens in a plane perpendicular to $\hat{f}$ vector).
The latter case is inspired by simulations of azobenzene polymers
\cite{azo_2005,azo_2011}. The approach can be termed as ``aided self-assembly'',
in contrast to the spontaneous
one. One should remark that the external field only promotes certain symmetry for
the molecular conformations but the molecules organise themselves into a bulk phase
by means of self-assembly.

The smectic-isotropic and columnar-isotropic transition temperatures are found to be
in the range of $490-500$~K and weakly dependent on
the number of attached chains $\Nch$ if $\Nch\leqslant  40$. This is attributed to the fact
that the mesogen-mesogen interactions are the same in all the cases. Therefore, to search for
ordered phases, the following steps are performed. First, the initial system is formed
by filling the simulation box randomly by LC macromolecules with $\Nch$ chains directed	
radially out of a central sphere. Then, the short $NVT$ run is performed at $T=500$~K with the time step
of $2\fs$ to remedy the beads overlapping. After that, several aided self-assembly
runs of duration $20~\ns$ are performed at $T=520$~K (above the LC transition) with the
timestep of $20\fs$ in $NP_{xx}P_{yy}P_{zz}T$ ensemble (for more details on this ensemble,
see \cite{IN07}). The runs differ by the value of a reduced field strength chosen
from the interval of $f=[3;5]$ for the uniaxial field and $f=[-5;-3]$ for the planar
one. Finally, the following runs are performed (mostly at $T=450$~K, about $50$~K below
the LC transition) in which the external field is removed, to check
on the stability of each bulk phase. All these runs are performed at non-zero external
pressure, as far the system is mostly density driven (out of all the non-bonded
interactions, equations~(\ref{Uspsp})--(\ref{Uscsc}), only the mesogen-mesogen pair
potential has an attractive contribution). The pressure of $53\mathrm{atm}$ is found
to be quite adequate for this purpose, as was found in an earlier study \cite{ILW2010}.

\begin{figure}
\begin{center}
\includegraphics[clip,width=9cm]{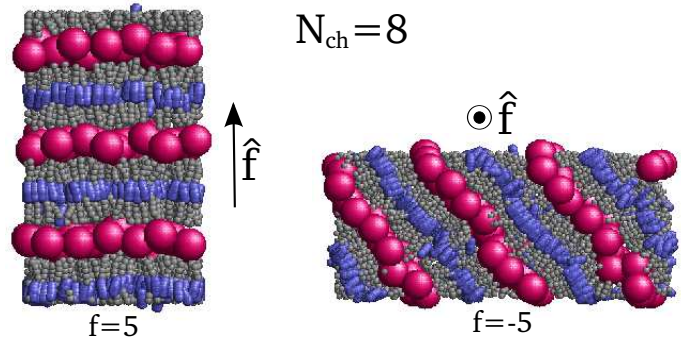}
\caption{\label{nch8_snap}(Color online) Results for an aided self-assembly of generic model
with $\Nch=8$ grafted chains. Left hand frame: uniaxial aiding field, right hand frame:
planar aiding field, field direction $\hat{f}$ is shown as arrow (points
towards the reader in the right hand frame). Note that the same lamellar smectic
phase is formed in both cases.}
\end{center}
\end{figure}

\begin{figure}[!t]
\begin{center}
\includegraphics[clip,width=4cm]{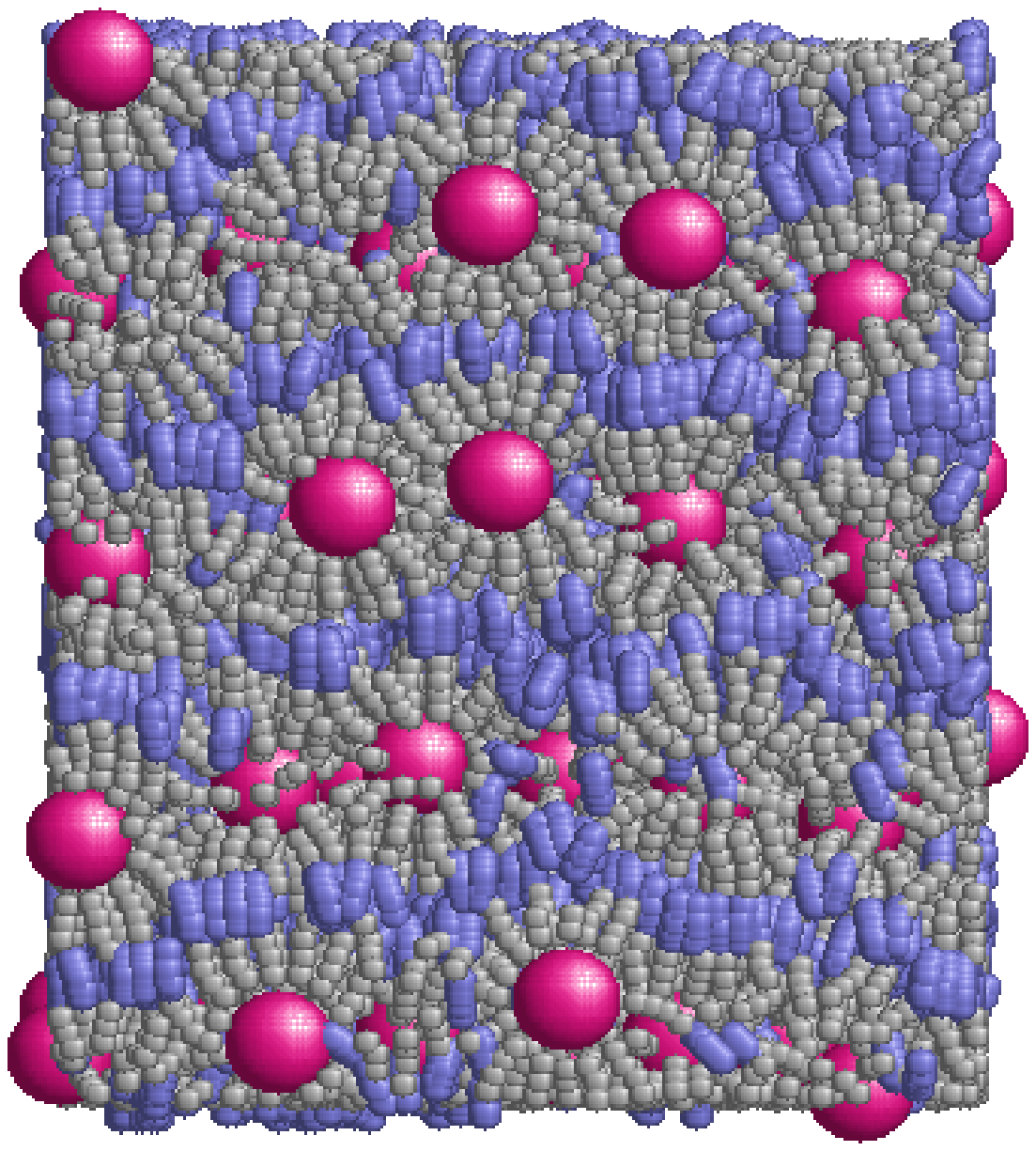}~~~
\includegraphics[clip,width=4.5cm]{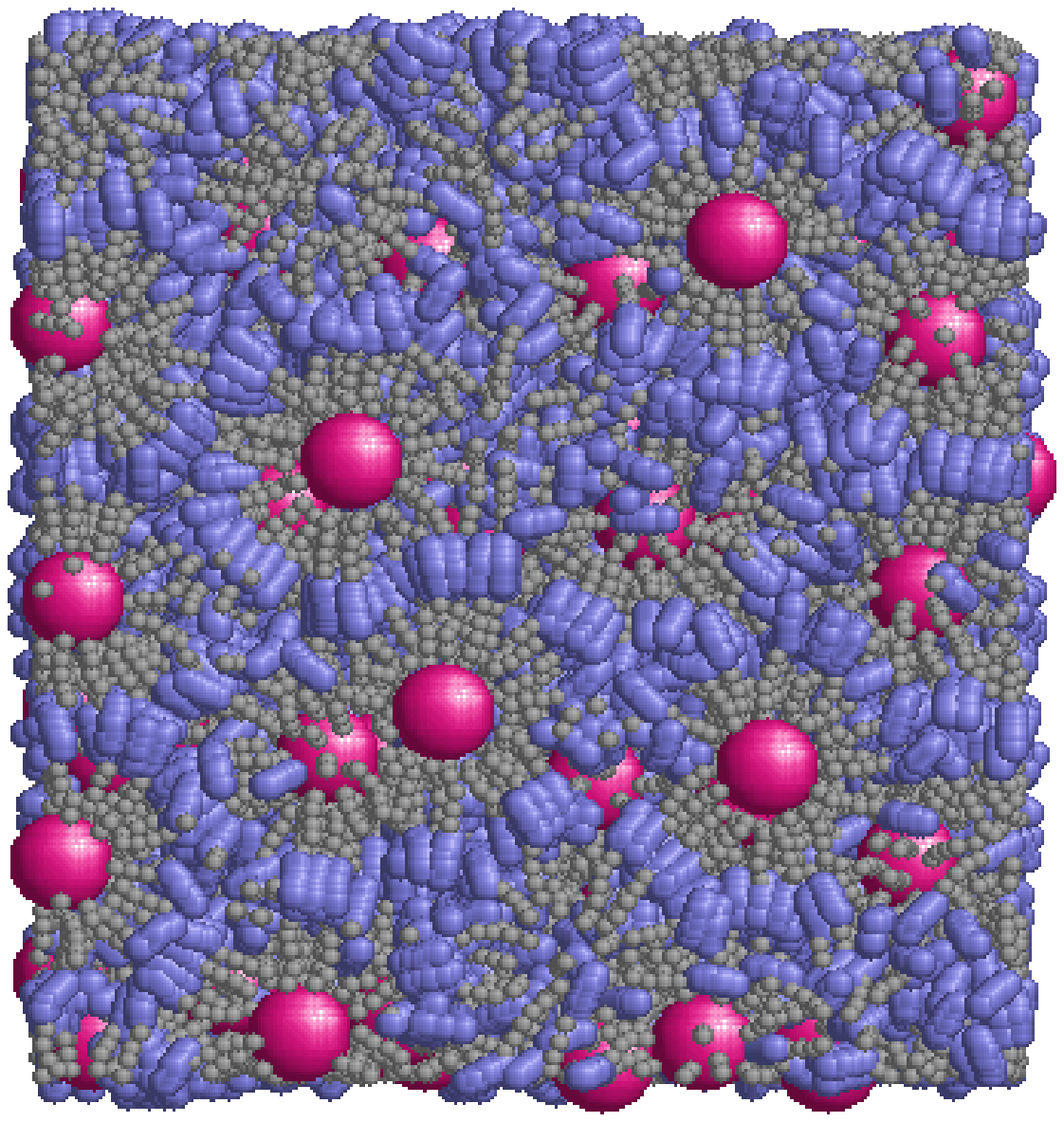}
\caption{\label{nch48}(Color online) Forced lamellar smectic phase for a generic model with $\Nch=48$
chains kept by means of uniaxial field (left hand frame) and break-up of this phase when the
field is switched ``off'' (right hand frame).}
\end{center}
\end{figure}

At the lower end of $\Nch$ values, the rod-like molecular conformation and bulk lamellar
smectic phase are expected. The self-assembly of this phase is aided by an uniaxial
field with $f>0$ at $T=520$~K. Nevertheless, for the sake of comparison, I also performed
runs for $f<0$ (attempting to force a discotic conformation).
In both cases, $\hat{f}$ is oriented along $Z$ axis and the runs of $10~\ns$  duration
are performed. After that, the field is removed and the system is equilibrated for another
$20~\ns$ at $T=450$~K. Remarkably, the same lamellar
smectic morphology is obtained in both cases (of $f>0$ and $f<0$), the layers only
differ in their arrangement with respect to the spatial axes (see, figure~\ref{nch8_snap}).
In particular, at $f=5$ the long axes of molecular rods are directed along $Z$ axis,
whereas at $f=-5$ they are confined within $XY$ planes.
In the latter case, the quasi-2D spontaneous self-assembly occurs inside these planes resulting
in the formation of the smectic layers. In both simulations with uniaxial and planar
fields, the rod-like conformation is observed only (the histograms will be provided
below), which says in favour of the aiding field approach. Indeed, the symmetry of
the field is not capable of forcing a certain conformation to occur (in this case -- a discotic
one), if it is not a native one for a given value of $\Nch$. The same scenario holds
for at least $\Nch=16$ attached chains, and in all these cases the lamellar smectic
phase is observed only. At the range of values of $\Nch=24-40$, the model displays
conformational bistability, discussed earlier in \cite{ILW2010}. In this case,
the symmetry of the aiding field acts as a conformation switcher. The largest number of chains at which the
smectic phase is observed is $40$, higher than the close-packing estimate
for the ``slim rod'' model (see, previous section) $N'=30$, thus, indicating a ``swollen
rod'' conformation. At a larger number of chains, $\Nch=48$, the lamellar smectic
phase can be forced by the field $f=5$, but it turns out to be unstable if the field is
removed and the temperature reduced to $450$~K (see, figure~\ref{nch48}).
Preliminary runs, performed for $\Nch=48$ in a temperature range of $T=[300,500]$~K,
indicate that the smectic-isotropic transition temperature in this case is much lower
than for the case of $\Nch=32$, namely $T\sim 400$~K {\it vs} $T\sim 490$~K,
respectively. These effects will be covered in detail in a separate study.

\begin{figure}[!b]
\begin{center}
\includegraphics[clip,width=4.5cm]{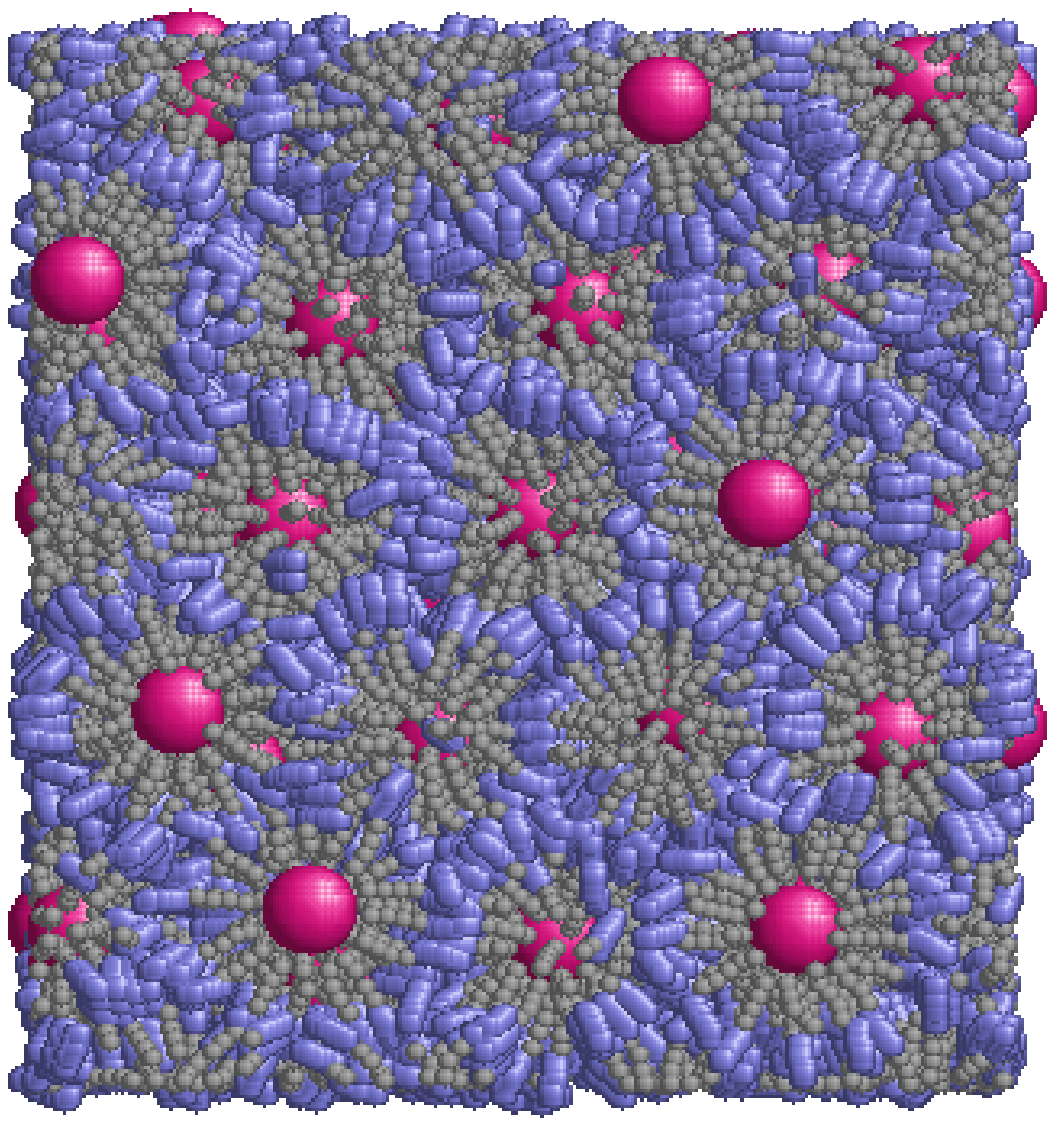}~~~
\includegraphics[clip,width=5cm]{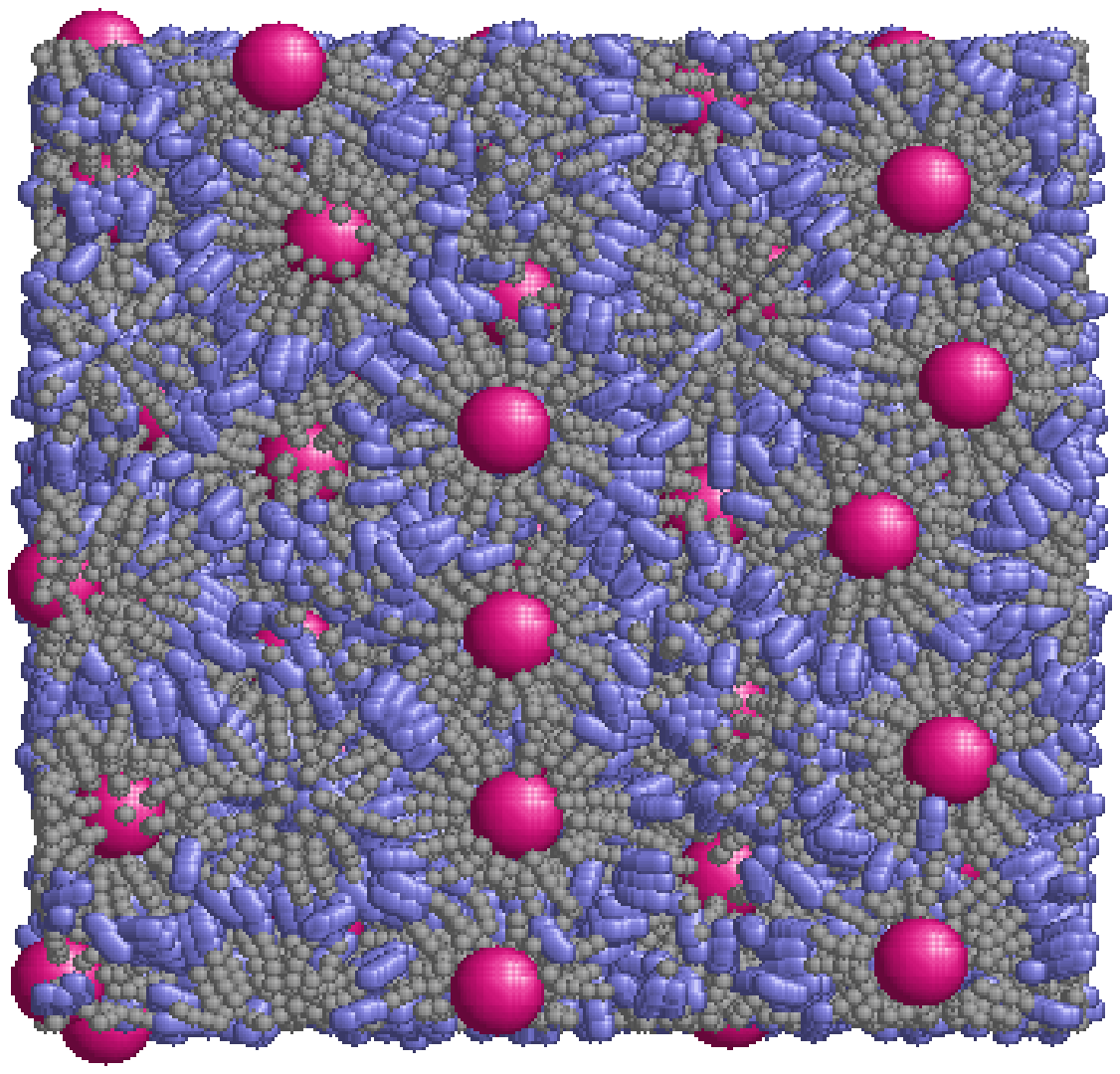}
\caption{\label{nch64}(Color online) Two views showing the symmetry of the cubic phase obtained
as the result of either spontaneous or aided with planar field self-assembly of
generic model with $\Nch=64$ chains. The image on the right resembles columnar structure
being swollen due to the change of molecular conformations from disc to a sphere.}
\end{center}
\end{figure}

The application of the planar field with $f<0$ induces a disc-like conformation and
aids self-assembly of a defect-free hexagonally packed columnar phase for
$\Nch=24-48$, including the case of $\Nch=32$ discussed in detail in reference
\cite{ILW2010}. The properties of
this phase and the snapshots are to be found there and are not repeated
here. At $\Nch\sim 56-64$, the discotic conformation transforms
into a spherulitic and, as a result, the cubic phase is formed (see, figure~\ref{nch64}).
Two views of the cubic phase are shown in this figure, and on the r.h.s. one may
identify the structure of swollen columns of the former columnar phase.
The interval of stability for the disc-like conformation in terms of $\Nch$ is
not spanning up to the value predicted by close packing of the grafting points,
$N^{*}\approx 55$, indicating not tightly packed discs.

\begin{figure}
\begin{center}
\includegraphics[clip,width=4cm]{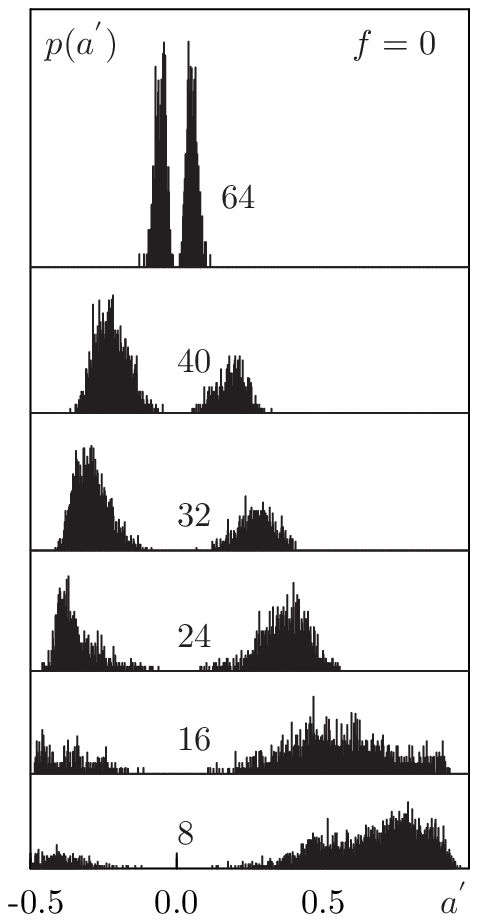}
\includegraphics[clip,width=4cm]{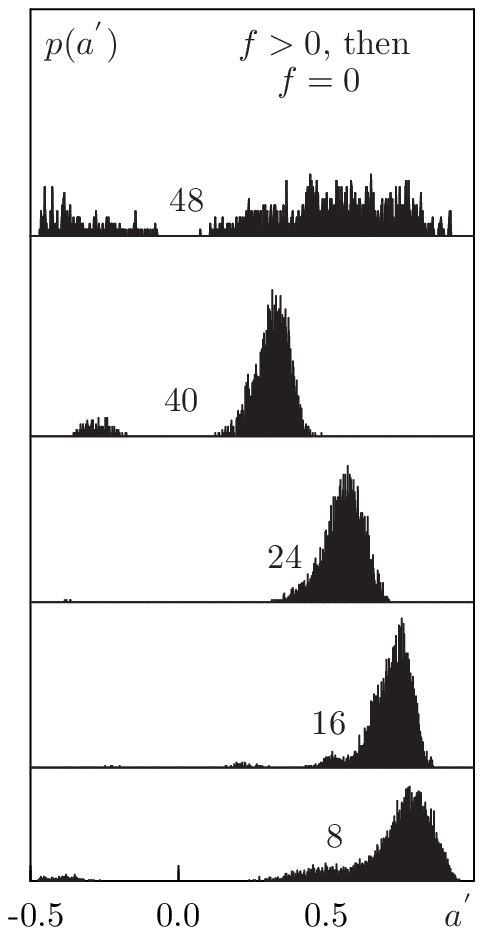}
\includegraphics[clip,width=4cm]{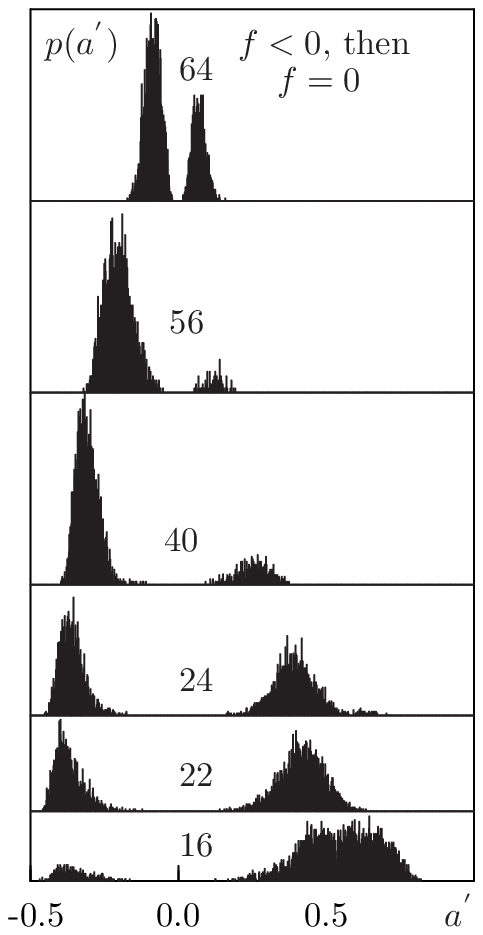}
\caption{\label{hist}Histograms for the distributions of molecular asphericity $p(a')$
(see text for explanations) shown for a spontaneous self-assembly (left hand image), uniaxial field
aided self-assembly (middle image) and planar field aided self-assembly (right hand image).
Only characteristic $\Nch$ cases are shown in each case.}
\end{center}
\end{figure}

Let me switch now to the quantitative analysis of conformations in the observed
bulk phases. To do so I split the system into rods and discs and build
histograms for asphericity of their conformations. First of all, the components
of gyration tensor are evaluated for each $k$-th molecule:
\begin{equation}\label{Gtens}
G_{\alpha\beta}^{[k]} = \frac{1}{N^{[k]}}\sum_{i=1}^{N^{[k]}}
   \left(r_{i,\alpha}^{[k]}-R_\alpha^{[k]}\right)\left(r_{i,\beta}^{[k]}-R_\beta^{[k]}\right),\qquad
\vec{R}^{[k]}=\frac{1}{N^{[k]}}\sum_{i=1}^{N^{[k]}}\vec{r_{i}}^{[k]},
\end{equation}
where $N^{[k]}$ particle centers with coordinates $r_{i,\alpha}^{[k]}$ are taken
into account, $R_\alpha^{[k]}$ is the molecular center of mass, $\alpha$,
$\beta$ denote Cartesian axes. To account for an extended shape of mesogens, each
is replaced by a line of four centers. The eigenvalues of gyration tensor,
$\lambda^{[k]}_{\mathrm{max}}$\,, $\lambda^{[k]}_{\mathrm{med}}$ and
$\lambda^{[k]}_{\mathrm{min}}$ (where the indices denote maximum,
medium and minimum value, respectively) are evaluated next. These are used to
introduce molecular ``roddicity'' (always positive):
\begin{equation}\label{ar}
a_{\mathrm{r}}^{[k]} = \left[\lambda^{[k]}_{\mathrm{max}} -
  \frac12(\lambda^{[k]}_{\mathrm{med}}+\lambda^{[k]}_{\mathrm{min}})\right][R^{[k]}_g]^{-2}
\end{equation}
and molecular ``discoticity'' (always negative):
\begin{equation}\label{ad}
a_{\mathrm{d}}^{[k]} = \left[\lambda^{[k]}_{\mathrm{min}} -
 \frac12(\lambda^{[k]}_{\mathrm{med}}+\lambda^{[k]}_{\mathrm{max}})\right][R^{[k]}_g]^{-2},
\end{equation}
for each $k$th molecule. Here, $[R^{[k]}_g]^2=\lambda^{[k]}_{\mathrm{max}}+
\lambda^{[k]}_{\mathrm{med}}+ \lambda^{[k]}_{\mathrm{min}}$ is squared radius of
gyration. If, for a given $k$, the ``roddicity'' prevails, $|a_{\mathrm{r}}^{[k]}|>|a_{\mathrm{d}}^{[k]}|$,
then it is classified as a rod and its asphericity is set to $a'=a_{\mathrm{r}}^{[k]}$,
otherwise the molecule is classified as a disc with its asphericity set to $a'=a_{\mathrm{d}}^{[k]}$.
As a result, the system splits into rods and discs subsystems, with their fractions $f_{\mathrm{r}}$
and $f_{\mathrm{d}}$, respectively. The histograms for $a'$ distribution $p(a')$ are built over all
the molecules in the system averaged over time trajectory. These are shown in figure~\ref{hist}
for some characteristic values of $\Nch$ in each case of spontaneous self-assembly (left hand frame),
and self-assembly aided by an uniaxial (middle frame) and planar (right hand frame) fields.
I did not include the histograms for $\Nch=12$ and $20$ (see, figure~\ref{nch_f=0}) into left
hand frame, as the self-assembly in these two cases were rather atypical as compared with
other cases of spontaneous self-assembly (these follow the route similar to an aided self-assembly).

The distributions of discoticity and roddicity are conveniently separated on these plots
as far as the former is negative and the latter is positive, the values close to zero indicate
spherulitic conformations. One can also see the relative weight of rod- and disc-like
conformations via the height of each wing, as well and the breadth of each distribution.
One may make the following observations from the histograms shown in figure~\ref{hist}.
In the case of a polydomain phase, as a result of spontaneous self-assembly (left hand frame),
rods and discs always coexist and the distributions of their asphericities are rather
broad. With an increase of $\Nch$, two maxima gradually merge into a spherulitic
shape from both sides of $a'=0$ (at about $\Nch=64$ and higher). The histograms
for the field-aided self-assembly are essentially narrower. In the case of uniaxial field
(middle frame) the discotic conformations are completely eliminated (except the case
of $\Nch=48$ where smectic phase is not observed any more), as these are incompatible
with the 1D symmetry of the aiding field. In the case of planar filed (right hand frame), which
has a 2D symmetry, the rod-like conformations are not eliminated and do appear within $XY$
plane, and are, in fact, the dominant ones at smaller values of $\Nch$ (as discussed
above for the case of $\Nch=8$, see figure~\ref{nch8_snap}). With an increase of $\Nch$
above $24$, the disc-like conformations dominate. Here, I would like to remind again that
the aiding field is switched ``on'' only at the beginning of each run, to promote the first
``kick'', followed by an extensive simulation with the field switched ``off''.
The comparison of histograms for spontaneous and aided self-assembly cases reveals
the effect of the aiding field in the form of conformation switching/enriching.
After the required conformations are enriched, the melt is capable of self-assembling
into an appropriate phase.

\begin{figure}
\begin{center}
\includegraphics[clip,width=5.5cm]{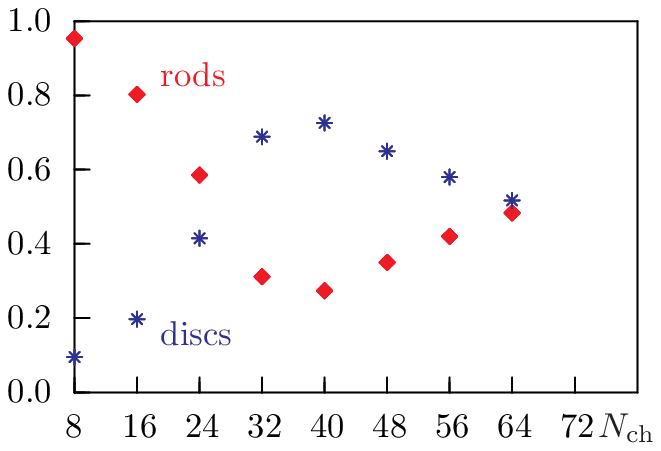}
\includegraphics[clip,width=5.5cm]{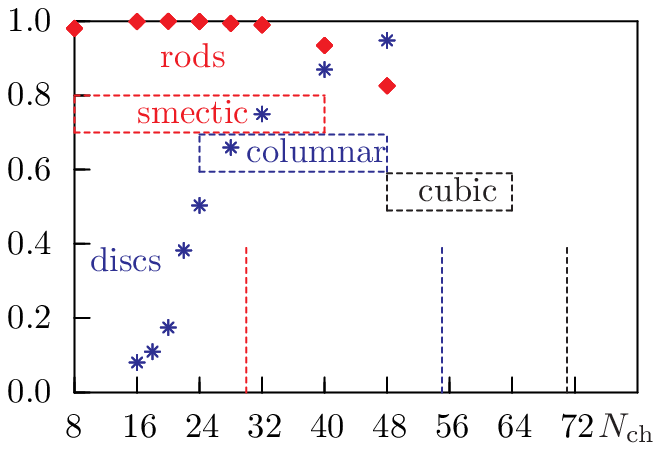}
\caption{\label{frac}(Color online) Fraction of rods and discs for spontaneous self-assembly (left frame,
$f_{\mathrm{r}}+f_{\mathrm{d}}=1$). The same properties are shown on the right but fraction of rods is shown
for uniaxial field aided runs and fraction of discs -- for planar field aided runs,
$f_{\mathrm{r}}+f_{\mathrm{d}}\neq 1$ in this case. The figure on r.h.s. shows also the approximate phase boundaries
for the smectic, columnar and cubic phases (dashed coloured horizontal text boxes)
and the optimal numbers for space-filling of rod, disc and sphere from geometry estimates
(vertical red, blue and black dashed lines, respectively).}
\end{center}
\end{figure}

The fractions of rods and discs, $f_{\mathrm{r}}$ and $f_{\mathrm{d}}$, as functions of $\Nch$ are displayed
in figure~\ref{frac} for various self-assembly runs. Left hand frame contains the data for a spontaneous
self-assembly and it indicates a broad region for a rod-disc coexistence at intermediate
values of $\Nch$. At $\Nch=64$, the system approaches a symmetric case with both
conformations transforming into a spherulitic shape. The right hand frame contains data for
$f_{\mathrm{r}}$ for uniaxial field aided self-assembly and data for $f_{\mathrm{d}}$ for planar field aided
self-assembly. Therefore, $f_{\mathrm{r}}+f_{\mathrm{d}}\neq 1$ as both are obtained for different cases.
One can see that the shapes of both curves are much steeper in this case as compared to
the left hand frame plot indicating once more the possibility to control the molecular
conformation by means of initial field of appropriate symmetry.

The comparison between the intervals with high molecular roddicity and discoticity with the
intervals of stability for the smectic and columnar phase (shown as coloured text boxes in
figure~\ref{frac}, on the right) shows their exact coincidence, thus indicating a strong
correlation between the average molecular shape and the symmetry of the bulk phase.
The space-filling geometrical estimates for slim rod and disc are 1.2--1.7 times larger
than the approximate mid-points of the respective intervals of stability of each phase.
Therefore, the real conformations considerably deviate from the ``slim''-like models.

\section{Conclusions}
\label{IV}

Computer simulations performed and discussed in this study provide some more
insight on a macromolecular self-assembly of liquid crystal colloids.
A generic model being used consists of a large central sphere and is modified on
its surface by grafted chains each terminated by a mesogen. The focus of
current study is on the role played by the surface density of chains on phase
diagram and typical molecular conformations.

Simple geometry estimates based on space-filling of macromolecule into a rod-like,
disc-like and spherulitic shape provided some reasonable starting point for the
relation between the number of grafted chains and equilibrium conformation.
Molecular dynamics simulations using soft interaction models repeat the
experimental evidence for the lamellar-columnar-cubic sequence of phases with
an increase of surface density. I found the model being conformationally bistable
at a wide range of surface density with the possibility to form either lamellar
smectic or columnar phase.

Conformational analysis is performed by introducing ``roddicity'' and
``discoticity'' of their shape and, therefore, sorting the molecules at each
time instance into rods and discs. The fraction of molecules in each subsystem
provides some preliminary information on the distribution of their conformations.
More details are provided by the histograms of their asphericity, these also
shed some light on a process of macromolecular self-assembly. In this respect,
the main obstacle in efficient self-assembly into a monodomain phases is
seen in a lack of control over the molecular conformations. In virtually
all the cases of surface density being considered, the rod- and disc-like
conformations coexist and have relatively broad distribution of their asphericity.

The problem can be partially remedied by an aided self-assembly used in this study.
It implies the use of an external field of certain symmetry (uniaxial, planar, etc.)
which acts on the mesogens orientations to promote specific conformations (rod-,
disc-like, etc.). When the bulk phase is formed, the field is removed and the
system is equilibrated at a desired temperature to check for the stability of thus
formed phase and to evaluate its properties. The problem of this approach is
a limited choice for the symmetry of the field and a bias towards specific phase which
should be known {\it a priori}. Another possible reason for, in general, poor
self-assembly of this particular model could be connected with the fact that
grafted chains are freely sliding on the large sphere resulting in broad
distributions for molecular asphericity and, as observed in some cases, an enhanced
microphase separation between large and small spheres.

This directs the following research in this area into refining the generic model
towards real systems and into developing some specific techniques to drive macromolecular
self-assembly.

\section*{Acknowledgements}

The paper is dedicated to the 70th birthday anniversary of professor
Myroslav Holovko, great scientist and teacher.

The author acknowledges participation in one of the workshops from the ``Mathematics
of Liquid Crystals'' series by INIMS (Cambridge, UK), 8--22 March 2013 and benefited
from exchange visits in the frames of EU Grant No. PIRSES 268498.

\ukrainianpart

\title{Взаємозв'язок між поверхневою густиною рідкокристалічної макромолекули та симетрією її саморганізованої фази: дослідження за допомогою методу огрубленої молекулярної динаміки}

\author{Я.М.~Ільницький}

\address{Інститут фізики конденсованих систем НАН України, вул. Свєнціцького, 1, 79011 Львів, Україна}

\makeukrtitle

\begin{abstract}
Розглянуто узагальнену модель, придатну для опису об'ємного впорядкування рідкокристалічних (РК) макромолекул (наприклад, РК дендримерів; наночастинок золота, модифікованих полімерними ланцюжками із кінцевими РК групами тощо). Дослідження концентрується на взаємозв'язку між кількістю приєднаних ланцюжків $\Nch$ та симетрією впорядкованої фази. Використовуючи прості геометричні обчислення спочатку оцінено інтервали стабільності для стержне-, диско- та сферо-подібних молекулярних конформацій залежно від $\Nch$. Далі виконано моделювання за допомогою молекулярної динаміки для спонтанного та керованого самовпорядкування РК макромолекул в об'ємні фази. Під час спонтанного самовпорядкування шляхом аналізу гістограм для молекулярної асферичності виявлено співіснування стержне- та диско-подібних конформацій в широкому інтервалі $\Nch$, що перешкоджає формуванню бездефектних структур. Використання одновісного або планарного керуючих полів суттєво покращує самовпорядкування відповідних монодоменних фаз шляхом селекції конформацій з відповідною симетрією. Сильна залежність між формою молекули та симетрією фази, яка спостерігається експериментально, також виявляється і при моделюванні -- через співпадіння інтервалів стабільності відповідних конформацій та об'ємних фаз.

\keywords макромолекули, рідкі кристали, самовпорядкування, молекулярна динаміка
\end{abstract}


\begin{thebibliography}{99}
\bibitem{Riess2003}
 Riess~G.,  Prog. Polym. Sci., 2003, \textbf{28}, 1107; \doi{10.1016/S0079-6700(03)00015-7}.
\bibitem{Olsen2008}
 Olsen~B.D.,   Segalman~R.A.,  Mat. Sci. Eng. R, 2008, \textbf{62}, 37; \doi{10.1016/j.mser.2008.04.001}.
\bibitem{Zeng1997}
 Zeng~F.,   Zimmerman~S.C.,  Chem. Rev., 1997, \textbf{97}, 1681; \doi{10.1021/cr9603892}.
\bibitem{Discher2002}
 Discher~D.E.,   Eisenberg~A.,  Science, 2002, \textbf{297}, 967; \doi{10.1126/science.1074972}.
\bibitem{Gittins2003}
 Gittins~P.J.,   Twyman~L.J.,  Supramol. Chem., 2003, \textbf{15}, 5; \doi{10.1080/1061027031000073199}.
\bibitem{Percec2004}
 Percec~V.,   Mitchell~C.M.,  Cho~W.-D.,   Uchida~S.,
 Glodde~M.,   Ungar~G.,   Zeng~X.,   Liu~Y., Balagurusamy~V.S.K.,
 Heiney~P.A.,  J. Am. Chem. Soc., 2004, \textbf{126}, 6078; \doi{10.1021/ja049846j}.
\bibitem{Tsc01b}
 Tschierske~C.,  J. Mater. Chem., 2001, \textbf{11}, 2647; \doi{10.1039/b102914m}.
\bibitem{SG05}
 Saez~I.M.,   Goodby~J.W.,  J. Mater. Chem., 2005, \textbf{15}, 26; \doi{10.1039/b413416h}.
\bibitem{Tschierske2007}
 Tschierske~C.,  Chem. Soc. Rev., 2007, \textbf{36}, 1930; \doi{10.1039/b615517k}.
\bibitem{Saez2008}
 Saez~I.M.,   Goodby~J.W., Struct. Bond., Liquid Crystalline Functional Assemblies and Their Supramolecular Structures, 2008, \textbf{128}, 1; \doi{10.1007/430_2007_077}.
\bibitem{Draper2011}
 Draper~M.,   Saez~I.M.,   Cowling~S.J.,   Gai~P.,   Heinrich~B.,   Donnio~B.,   Guillon~D.,
 Goodby~J.W.,  Adv. Funct. Mater., 2011, \textbf{21}, 1260; \doi{10.1002/adfm.201001606}.
\bibitem{Kumar2011}
 Bisoyi~H.K.,   Kumar~S.,  Chem. Soc. Rev., 2011, \textbf{40}, 306; \doi{10.1039/b901793n}.
\bibitem{PBS+00}
 Ponomarenko~S.A.,   Boiko~N.I.,   Shibaev~V.P.,   Richardson~R.,   Whitehouse~I.,
 Rebrov~E.,   Muzafarov~A.,  Macromolecules, 2000, \textbf{33}, 5549; \doi{10.1021/ma0001032}.
\bibitem{Agina2007}
 Agina~E.V.,   Boiko~N.I.,   Richardson~R.M.,   Ostrovskii~B.I.,
 Shibaev~V.P.,   Rebrov~E.A.,   Muzafarov~A.M.,  Polym. Sci. Ser. A, 2007, \textbf{49}, 412; \doi{10.1134/S0965545X07040086}.
\bibitem{Wojcik2009}
 Wojcik~M.,   Lewandowski~W.,   Matraszek~J.,   Mieczkowski~J.,   Borysiuk~J.,
 Pociecha~D.,   Gorecka~E.,  Angew. Chem. Int. Edit., 2009, \textbf{48}, 5167; \doi{10.1002/anie.200901206}.
\bibitem{Wojcik2010}
 Wojcik~M.,   Kolpaczynska~M.,   Pociecha~D.,   Mieczkowski~J.,   Gorecka~E.,
Soft Matter, 2010, \textbf{6}, 5397; \\ \doi{10.1039/c0sm00539h}.
\bibitem{Wojcik2011}
 Wojcik~M.,   Gora~M.,   Mieczkowski~J.,   Romiszewski~J.,   Gorecka~E.,
Soft Matter, 2011, \textbf{7}, 10561;\\ \doi{10.1039/c1sm06436c}.
\bibitem{WIS03}
 Wilson~M.R.,   Ilnytskyi~J.M.,   Stimson~L.M.,  J. Chem. Phys., 2003, \textbf{119},
  3509; \doi{10.1063/1.1588292}.
\bibitem{Balabaev_old}
 Mazo~M.A.,   Shamaev~M.Yu.,   Balabaev~N.K.,   Darinskii~A.A.,   Neelov~I.M.,
Phys. Chem. Chem. Phys., 2004, \textbf{6}, 1285; \doi{10.1039/b311131h}.
\bibitem{Balabaev_new}
 Markelov~D.A.,   Mazo~M.A.,   Balabaev~N.K.,   Gotlib~Yu.Ya.,  Polym. Sci. Ser. A, 2013,
\textbf{55}, 53; \\\doi{10.1134/S0965545X13010045}.
\bibitem{HWS05}
 Hughes~Z.E.,   Wilson~M.R.,   Stimson~L.M.,  Soft Matter, 2005, \textbf{1}, 436; \doi{10.1039/b511082c}.
\bibitem{ILW2010}
 Ilnytskyi~J.M.,   Lintuvuori~J.,   Wilson~M.R.,  Condens. Matter Phys., 2010, \textbf{13},
33001; \doi{10.5488/CMP.13.33001}.
\bibitem{OZ2013}
 Orlandi~S.,   Zannoni~C.,  Mol. Cryst. Liq. Cryst., 2013, \textbf{573}, 1; \doi{10.1080/15421406.2012.763213}.
\bibitem{LW08}
 Lintuvuori~J.S.,   Wilson~M.R.,  J. Chem. Phys., 2008, \textbf{128}, 044906; \doi{10.1063/1.2825292}.
\bibitem{EIW01}
 Earl~D.J.,   Ilnytskyi~J.,   Wilson~M.R.,  Mol. Phys., 2001, \textbf{99}, 1719; \doi{10.1080/00268970110069551}.
\bibitem{IW_2000} Ilnytskyi~J., Wilson~M.R., Comput. Phys. Commun., 2001, \textbf{134}, 23; \doi{10.1016/S0010-4655(00)00187-9}.
\bibitem{IW_2001} Ilnytskyi~J., Wilson~M.R., Comput. Phys. Commun., 2002, \textbf{148}, 43; \doi{10.1016/S0010-4655(02)00467-8}.
\bibitem{GM98}
 Groot~R.D.,   Madden~T.J.,  J. Chem. Phys., 1998, \textbf{108}, 8713; \doi{10.1063/1.476300}.
\bibitem{IlnPatsHol2008}
 Ilnytskyi~J.,   Patsahan~T.,   Holovko~M.,   Krouskop~P.,   Makowski~M.,  Macromolecules, 2008,
  \textbf{41}, 9904; \\ \doi{10.1021/ma801045z}.
\bibitem{Bates2009}
 Bates~M.,   Walker~M.,  Soft Matter, 2009, \textbf{5}, 346; \doi{10.1039/b813015a}.
\bibitem{Bates2009a}
 Bates~M.A.,   Walker~M.,  Phys. Chem. Chem. Phys., 2009, \textbf{11}, 1893; \doi{10.1039/b818926a}.
\bibitem{LW09}
 Lintuvuori~J.S.,   Wilson~M.R.,  Phys. Chem. Chem. Phys., 2009, \textbf{11}, 2116; \doi{10.1039/b818616b}.
\bibitem{azo_2005}
Ilnytskyi~J.M., Neher~D., Saphiannikova~M., Condens. Matter Phys., 2006, \textbf{9}, 87; \doi{10.5488/CMP.9.1.87}.
\bibitem{azo_2011}
Ilnytskyi~J.M., Neher~D., Saphiannikova~M.,  J. Chem. Phys., 2011, \textbf{135}, 044901; \doi{10.1063/1.3614499}.
\bibitem{IN07}
 Ilnytskyi~J.M.,   Neher~D.,  J. Chem. Phys., 2007, \textbf{126}, 174905; \doi{10.1063/1.2712438}.
\end{thebibliography}
\end{document}